\documentclass[10pt]{article}

\usepackage{amsmath}
\usepackage{amssymb}
\usepackage{longtable}

\usepackage{graphicx}
\usepackage{caption}
\usepackage{epstopdf} 
\usepackage{multirow}

\usepackage{dcolumn}

\usepackage{cite}
\usepackage{color} 

\usepackage{subfig}

\makeatletter
\renewcommand{\@biblabel}[1]{\quad#1.}
\makeatother

\date{}

\begin{document}

\begin{flushleft}
{\Large
\textbf{Cities through the Prism of People's Spending Behavior}
}
\\
\bigskip
{\large
Stanislav Sobolevsky$^{1,*,**}$, 
Izabela Sitko$^{2,**}$, 
Remi Tachet des Combes$^{1}$,
Bartosz Hawelka$^{2}$, 
Juan Murillo Arias$^{3}$,
Carlo Ratti$^{2}$}
\\
\bigskip
{1} Senseable City Lab,  Massachusetts Institute of Technology, Cambridge, MA, United States of America
\\
{2} Department of Geoinformatics - Z\_GIS, University of Salzburg, Salzburg, Austria
\\
{3} New Technologies, BBVA, Madrid, Spain
\\
$*$ Correspondence: E-mail: stanly@mit.edu
\\
$**$ These authors contributed equally to this work.
\end{flushleft}

\section*{Abstract}
Scientific studies of society increasingly rely on digital traces produced by various aspects of human activity. In this paper, we use a relatively unexplored source of data--anonymized records of bank card transactions collected in Spain by 
a big European bank-- in order to propose a new classification scheme of cities based on the economic behavior of their residents.
First, we study how individual spending behavior is qualitatively and quantitatively affected by various factors such as customer’s age, gender, and size of his/her home city. We show that, similar to other socioeconomic urban quantities, individual spending activity exhibits a statistically significant superlinear scaling with city size. With respect to the general trends, we quantify the distinctive signature of each city in terms of residents' spending behavior, independently from the effects of scale and demographic heterogeneity.
Based on the comparison of city signatures, we build a novel classification of cities across Spain in three categories. That classification is, with few exceptions, stable over different ways of city definition and connects with a meaningful socioeconomic interpretation. 
Furthermore, it appears to be related with the ability of cities to attract foreign visitors, which is a particularly remarkable finding given that the classification was based exclusively on the behavioral patterns of city residents. This highlights the far-reaching applicability of the presented classification approach and its ability to discover patterns that go beyond the quantities directly involved in it.

\section*{Introduction}
Laws and regularities in human behavior have been the subject of intense research for several decades. 
In the age of ubiquitous digital media, different aspects of human activity are being increasingly analyzed by means of 
their digital footprints, such as mobile call records \cite{ girardin2008digital, gonzalez2008uih, quercia2010rse, sobolevsky2013delineating, amini2014impact}, vehicle GPS traces \cite{santi2014}, social media activity \cite{szell2013, frank2013happiness, hawelka2014, paldino2015} or smart card usage \cite{bagchi2005, lathia2012}.
The wide popularity of debit and credit cards, which are increasingly replacing cash spending, suggests the appearance of yet another source of valuable information for scholars across a wide range of disciplines. The extensive transactions data set, collected by banks and other providers of payment systems, allows a glimpse into the daily activities of large numbers of individuals. While the spatiotemporal granularity of these data may be sparser compared with previously explored sources, such as call detail records (CDRs), incorporated contextual information such as spending category, amount, and place enables the analysis not only of movement patterns but also of their semantics--i.e., the context of human activity.  
Furthermore, demographic profiles of bank customers provide an additional important layer that could explain such activity. 

In urban studies, the analysis of aggregated bank card data may provide novel insights in the description and comparison 
of the economic dimension of cities, in a way that goes beyond other types of digital records used in the past to investigate urban \cite{louail2014citystruct} and regional structure \cite{ratti2010}, land use \cite{grauwin2014towards, pei2014}, mobility \cite{noulas2012tale, kung2014exploring}, or well-being \cite{lathia2012}. Adequate and reliable metrics are of a primary importance for a city management, especially in the context of the increasingly competitive global economy \cite{arribasbell2013}. In this paper, we propose to look at the socioeconomic conditions of urban areas through the lens of the spending activity of their inhabitants, based on the dataset that ensures uniform and up-to-date description of a multitude of cities and the approach that compensates for their various sizes and characters. 

Previous studies of individual economic activity were mostly based on field studies \cite{lloyd54si}, questionnaires \cite{childers2001}, and surveys \cite{dholakia1999}. Direct analysis of individual bank card records has not been extensive so far. As card transaction data are highly sensitive including a lot of private information and therefore requires a lot of effort in anonymizing the data appropriately, their access has been so far highly restricted. Hitherto applications have mostly been focused on the card system itself \cite{chan1999frauddetection, rysman2007}, rather than on the associated human behavior. 
More recently, Krumme et al. \cite{krumme2013patterns} employed this new type of data to uncover the predictability of spending choices, and their relationship to wealth. The analysis of bank card data was also carried out in the field of regional delineation \cite{sobolevsky2014money}, human mobility \cite{lenormand_bcards_2014}, as well as the assessment of city attractiveness for different groups of customers \cite{sobolevsky2014mining, sobolevsky2015attract}. However, a comprehensive analysis of human spending behavior in cities has not yet been performed--such is the scope of the present study.

Shopping patterns of individuals were found to depend on the demographic factors such as gender, age, education, occupation, or income \cite{dholakia1999, bhantagar2000, hui2007, lenormand_bcards_2014}. However, the character of this relation on the propensity of bank card usage is reported differently across different studies.
Regarding gender, some results indicate an increased likelihood of bank cards usage among women \cite{hayhoe2000, borzekowski2008}, while others point to their preference for checks over cash or cards \cite{bounie2006}. Women are further reported to spend more money in the higher number of transactions than men~\cite{lenormand_bcards_2014}.
Age is reported to either lower the probability of card usage \cite{borzekowski2008}, have no significant effect \cite{bounie2006}, or finally decrease or increase spending activity depending on gender~\cite{lenormand_bcards_2014}. Given those discrepancies, as well as the possibly different character of the analyzed cities, we begin this study with a detailed analysis of the impact of a demography on individual bank card transactions and normalize the aggregated spending profiles of cities accordingly.  

Another important context of human economic activity is geographic location. In the case of urban customers, this notably concerns the city of residence and its size, described in terms of population as the major characteristic. Due to agglomeration effects and intensified human interactions, a variety of urban processes have been shown to vary with the number of inhabitants in the form of the scaling laws \cite{batty2008size, bettencourt2007growth, brockmann2006scaling, schlapfer2012scaling}. While urban infrastructure dimensions (e.g., total road surface) reveal a sublinear relation to city size, socioeconomic quantities (e.g., gross metropolitan product, crime rate, patenting, and human interactions) usually increase in a superlinear manner \cite{bettencourt2013origins}. One can expect that human spending in urban areas holds similar property, which is one of the hypothesis that will be tested below. 

In this paper, individual purchase activity via bank card records is explored in order to discover collective patterns of economic activity. Based on those, we propose a novel approach to compare and classify Spanish cities in terms of the spending behavior of their inhabitants. The results are presented within three main sections. 

We start by asking a broad range of questions regarding the fundamental factors that impact economic conduct. At first, we reveal the ways that demographic factors, such as age and gender, influence five representative quantities of individual bank card spending in Spain during the year 2011. In the further analysis, this step allows to look beyond the impact of a demographic heterogeneity between cities. In the next part, we investigate the impact of a city of residence, studying whether people from different places tend to spend their money in different ways. In line with previous studies on the scaling laws governing urban quantities, we examine the impact of city size on the economic activity of its inhabitants. 
Discovered trends give a good reference point for the expected collective behavior in a city with a given size. However, each particular city demonstrates its unique performance. Therefore in the final section, following the approach of \cite{bettencourt2010urbscaling}, we propose an index that measures the relative performance of cities based on the deviations of spending parameters from the general trends, indirect effects of population specificity taken into account. Such index allows for the comparison of cities of different population sizes and forms the basis for a novel scale-free classification of Spanish cities based on the economic behavior of the residents. The classification is given thorough geographic and socioeconomic interpretation, revealing meaningful patterns and different characters of urban areas in Spain. The most particular finding concerns the correspondence between the spending profile of cities and their ability to attract foreign visitors. Although the classification is constructed exclusively based on the behavior of residents, we found that certain city categories perform pretty differently in terms of their attractiveness. This highlights the far-reaching applicability of the proposed approach going beyond those quantities that are actually involved in it.

\section*{Materials and Methods}

\subsection*{Data set of bank card transactions}
Our study relies on the complete set of bank card transactions (both debit and credit) performed by the Spanish customers of Banco Bilbao Vizcaya Argentaria (BBVA) within the country in 2011.
The total number of active customers reaches around 4.5 M, who executed more than 178 M transactions, with a cumulative spending exceeding 10.3 billion euro. Due to the sensitive nature of bank records, they were anonymized  by BBVA prior to sharing, in accordance to all local privacy protection laws and regulations. Randomly generated IDs of customers are connected with certain demographic characteristics and an indication of a residence at the level of zip code, further aggregated into coarser spatial units. Each transaction is denoted with its value, a time stamp, a location of a point of sale where it was performed, and the business category it belonged to. The business classification includes 76 categories such as restaurants, gas stations, supermarkets or travels. 
In order to compensate for the inhomogeneous penetration of BBVA on the individual banking market in Spain, we normalize the activity of customers by the BBVA market share in the respective residence location (provided by the bank).
The raw data set is protected by a nondisclosure agreement and is not publicly available. However, certain aggregated data may be shared upon request for the purpose of findings validation.

\subsection*{Major characteristics of customers' spending behavior}
In order to characterize the spending behavior of customers, we consider
five basic parameters of bank card usage. Three of them are related to
the economic dimension of transactions:
\begin{itemize}
\item the activity of each customer, defined as the total number of
transactions performed during a year,
\item The average value of a single transaction
\item The spending diversity, measured by the number of distinct business
subcategories visited by a~customer in 2011
\end{itemize}
\noindent
Additionally, we introduce two characteristics of customers' mobility:
\begin{itemize} 
\item Distant mobility, measured as the percentage of transactions
executed over 200 km from home
\item Local mobility, measured as the average distance between the
customer's home location and the retail points (calculated based on
transactions made within 100 km from home)
\end{itemize}

Correct computation of four of the aforementioned quantities (all but activity) requires the customer to be using his bank card
frequently enough (e.g., there is no point in measuring the spending diversity or mobility
of someone who used a card only a couple of times). In the further analysis,
we thus only consider customers who performed at least 50
transactions in 2011 (which gives an average close to one transaction per week). 
Moreover, we restrict the analysis to customers active during the entire year i.e., 
those who performed at least one transaction during both the first and the last month of 2011. 
All five characteristics of spending behavior are further considered at the city scale--as an average value across the activity of residents.

\subsection*{Three levels of city definition}
For a spatial definition of Spanish cities we test three different types of units. The coarser city level consists of 24  Large Urban Zones (LUZs) as defined by the European Urban Audit Survey~\cite{eur_uaudit}. 
The intermediate level concerns 211 Conurbations (CONs) identified within the AUDES project (A\'reas Urbanas de Espan\~a~) \cite{audes}. 
For the finer spatial scale, we aggregate Administrative Cities of Urban Audit into 40 Functional Urban Areas (FUAs), so as to reflect metropolitan regions in agreement with the Study on Urban Functions of the European Spatial Planning Observation Network (ESPON)~\cite{espon2007}. Population and socioeconomic statistics for LUZ and FUA levels were obtained from Eurostat~\cite{eur_uaudit} and the National Statistics Institute of Spain \cite{ine}. Population figures for the CON level comes from the AUDES project.

\section*{Impact of demography on customer behavior}

\begin{figure}[h]
\vspace{0.3 cm}
\centering
\includegraphics{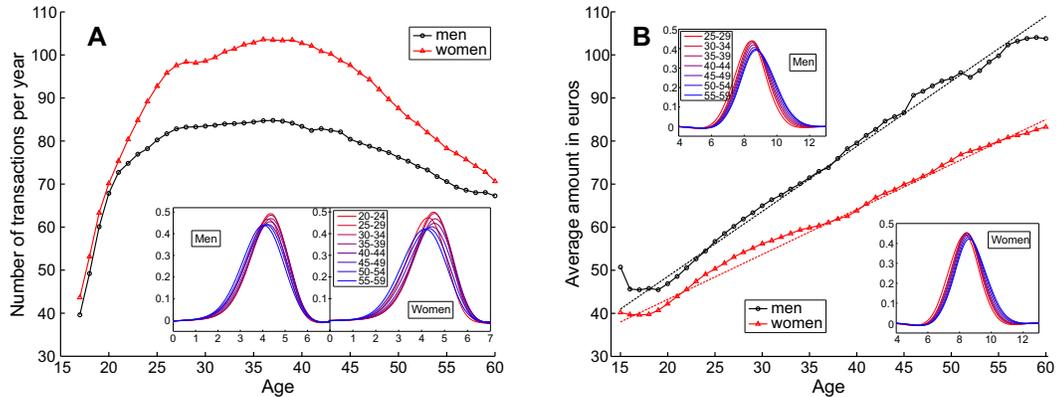}
\caption{{\bf Impact of age and gender on selected parameters of customers' spending behavior.}
 A: The average number of transactions per year. B: The average value
of a single purchase. Insets: Logarithmic distributions for both genders and different age groups.}
\label{fig::age_impact_activity}
\end{figure}

Among the primary factors one could think about to affect human  
economic behavior are age and gender \cite{bhantagar2000, dholakia1999, hui2007}. Sociodemographic characteristics were also demonstrated to affect human spending habits and mobility \cite{lenormand_bcards_2014}.
In this study, we explore how the distributions of the five measures of bank card usage (i.e., customer's activity, average value of a transaction, diversity of transactions, as well as distant and local mobility) change with customer age for both genders. 
We present results for respective parameters in Fig. \ref{fig::age_impact_activity}--\ref{fig::age_impact_mobility}. 
From a global perspective, one can observe that even though trends for both genders are substantially different quantitatively, in most cases, they exhibit remarkable similarities in their shape from a qualitative viewpoint. For instance, the number of transactions is usually higher for women, while the average value per transaction is higher for men, who seem to concentrate their economic activity more than women. Also, while the spending diversity of women customers is higher, their mobility is substantially lower on average. Nevertheless, the tendency for both men and women, as well as the important age thresholds where these tendencies change, appear to be strikingly similar. Looking beyond the simple average values of each characteristic i.e., analyzing statistical distributions, we again observe the steady and continuous impact of customers' demographics on their shape. Let us now take a closer look at the respective parameters.

\subsection*{Customer activity and average amount per transaction}

In Fig.~\ref{fig::age_impact_activity}A, the average number of
transactions per year is plotted against age, for both men (in black) and
women (in red). We also aggregated the data into age groups and plotted
the distribution of the transactions number (in log) for five-year
brackets. One can see
that customer activity increases rapidly between 18 and 30 years old, as
expected with the entry into the workforce.
It then reaches its peak and remains more or less constant for both genders till
40 years, before starting to steadily decrease. From an economic point of
view, it thus appears that people are most active during their
30s. Moreover, comparing the two curves shows that women make every
year on average 16.1\% more transactions; that number even goes up to 20.4\% when considering customers in the 25--45 age group. 

After the number of transactions, let us focus on their average amount (Fig.~\ref{fig::age_impact_activity}B). 
It is quite remarkable  
that this quantity grows with the customers' age 
in a nearly linear way, doubling between the youngest (18 years old) and 
the oldest (60 years old) customers. This effect could be intuitively explained by 
the ability of older people to spend more or to buy more expensive goods (they usually earn more money as their career develops).  
Nevertheless, it does not go in line with the pattern displayed by the
total amount of spending, a quick increase till 40 years old and then no
more variations (the actual graph can be found in S1 Fig.). Thus, it seems more natural 
to explain the steady increase of the average purchase value by a habit of 
concentrating purchases, making fewer transactions but buying more each time.

As to gender differences, the average amount per
transaction is smaller for women, who appear to spend more often, but for
smaller values. These two statements might be related to the following
demographic fact: between the age of 20 and 60 years old, 75\% of women
and  89\% of men are active \cite{ine}. And even though the situation is evolving
(the percentage of working women was only 65\% in 2005), one cannot help
but think that women remain more involved in domestic tasks, in particular
the essential shopping. Consequently, they would be bound to use their
bank card more frequently and for smaller amounts.

\subsection*{Spending diversity}

\begin{figure}[h!]
\centering
\includegraphics{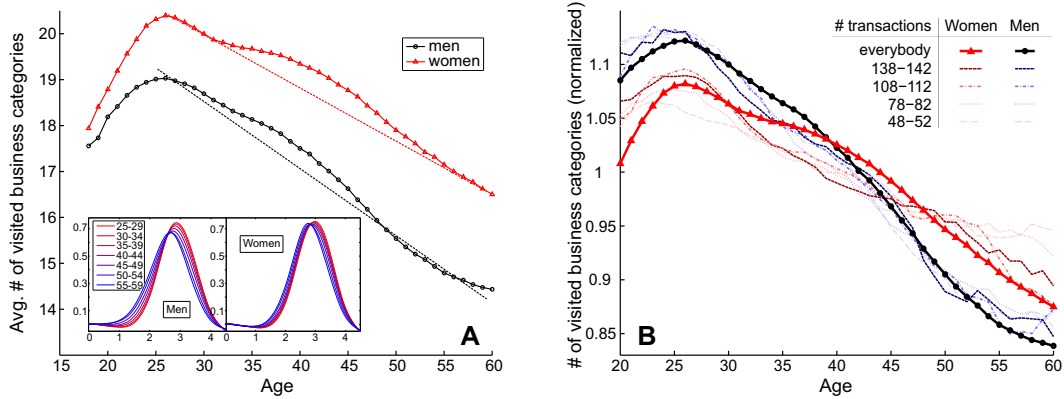}
\caption{{\bf Impact of age and gender on customers' spending diversity.}
A: Average spending diversity  against age for men and women. B:  Normalized number of visited business categories for the customers with different levels of spending activity.}
\label{fig::age_impact_diversity}
\end{figure}

Given the data at hand, we are also able to analyze a diversity of human economic behaviors based on a variety of places where people spend their money. In Fig.~\ref{fig::age_impact_diversity}, we correlate the average number of business categories visited over a year with customers' age and gender. 
After a rapid increase until the age of 27--28, the diversity of spending declines steadily in a linear way, showing that, while  people spend more money growing old, they also tend to spend it in fewer types of businesses. 
One can only wonder if the abrupt change in the trend occurring in the late 20s is related to the foundation of a family.

As far as gender is concerned, we notice that women tend to visit a larger number of business categories. This raises a question, however. In Fig.~\ref{fig::age_impact_diversity}A, the average was taken on every customer of a given age regardless of their total number of transactions, and we have seen that age greatly impacts the transaction activity (which in turn could impact diversity). To ensure that diversity measures are not biased by different activity levels at different ages, we group customers according to their level of activity (from 50 +/- 2 to 140 +/- 2) and plot in Fig.~\ref{fig::age_impact_diversity}B the normalized number of visited business categories for each group. We also plot the same quantity for the entire set of customers (the thick black and red lines). The graphs exhibit the pattern already seen in Fig.~\ref{fig::age_impact_diversity}A, which confirms our previous conclusions.

\subsection*{Customer's mobility}

While the home location of a customer is irrelevant to the aforementioned considerations, it becomes essential when studying human mobility. In the data, for each anonymized customer ID is given a zip code of the residency address. However, as far as each individual customer is concerned, the exactness of this formal zip code is questionable (e.g., students registered at their parents' home, people moving and not informing their bank, etc.). To get rid of that bias, we compute the fraction of transactions that took place in the daily accessible neighborhood of the reported home zip code and discard all customers for which said value is smaller than 60\%. They represent around 18\% of customers. 

\begin{figure}[h!]
\centering
\includegraphics{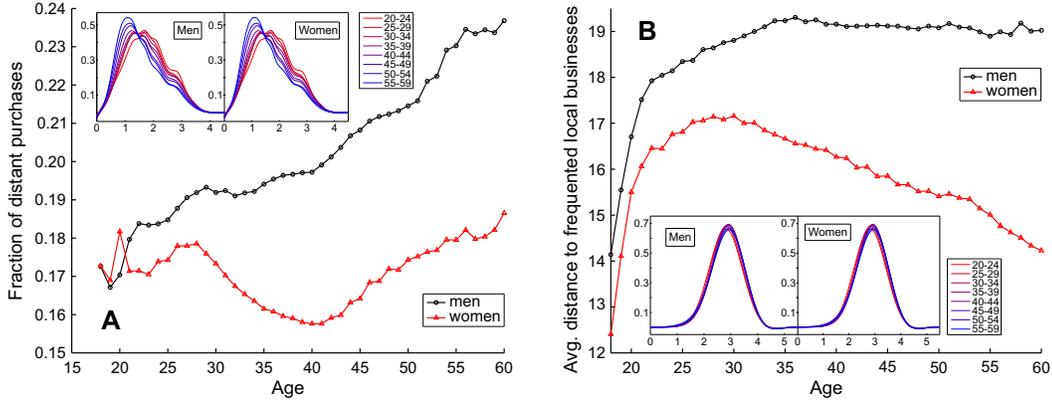}
\caption{{\bf Impact of age and gender on customer mobility.}
 A: Frequency of distant travels. B: The average distance to home of local transactions. Insets: Logarithmic distributions for both genders and different age groups.}
\label{fig::age_impact_mobility}
\end{figure}

In Fig.~\ref{fig::age_impact_mobility}, we plot the two parameters of customers' mobility (percentage of transactions performed more than 200 km away from home and average distance traveled to businesses less than 100 km from home) against age for both genders. The distant mobility (Fig.~\ref{fig::age_impact_mobility}A) is the first quantity displaying a big difference in trends between men and women under 40 years old. While for men, the fraction of distant purchases increases with age in a roughly linear way, distant mobility of women firsts stagnates, and then decreases until 40 years old, and finally starts to increase similarly to the curve observed for men. In a parallel way to the analysis of activity, one can think of societal explanations to cast light on such a difference. The average age for childbearing in Spain is 29.8 years old \cite{worldfactobook}, which strikingly corresponds to the change in the curve evolution. Regarding local mobility, an interesting pattern can be identified in Fig.~\ref{fig::age_impact_mobility}B. While local mobility of men remains nearly stable after 25--30 years old, with only a very slight tendency to decrease with age, women exhibit significant and stable decrease of the average distance to visited local retailers. They tend to shop closer to their home when growing older, which, together with the overall shorter distance of local purchases, well agrees with the findings of \cite{lenormand_bcards_2014}.

\section*{Bigger cities boost up spending activity}

It is well established that living in a bigger city boosts up many aspects of human life: intensity of interactions \cite{schlapfer2012scaling}, creativity \cite{bettencourt2010urbscaling}, economic efficiency (e.g., measured in GDP \cite{bettencourt2013origins}), as well as certain negative aspects such as crime \cite{bettencourt2010urbscaling}. In the following section, we examine whether this property holds true for the individual economic activity of city residents. To do so, average values of our five  bank card usage characteristics are quantified and their dependence on city size (expressed in terms of population) is analyzed. As the urban scaling laws were found sensitive to the selection of city boundaries \cite{arcaute2013citybound}, we test and compare three levels of city definition, namely 24 Large Urban Zones (LUZs), 40 Functional Urban Areas (FUAs), and 211 Conurbations (CONs).

In the previous section, we prove individual economic behavior to depend on customers' demography. It thus appears necessary to take into account possible variations of demographic profiles between different cities. As a matter of fact, age and gender vary quite significantly from one city to another. Among the 24 LUZs, for instance, the fraction of male customers varies between 47.5\% and 51\%, the average customer age goes from 41 to 48 years old, and the respective fractions of different age groups change up to a factor of 1.7. In order to correct for that demographic heterogeneity, we normalize each of the city characteristics by their expected value (computed using the demographic composition of the city and the average parameters estimated on the entire set of customers; for more details, see S1 Text). 

\begin{figure}[h]
\centering
\includegraphics{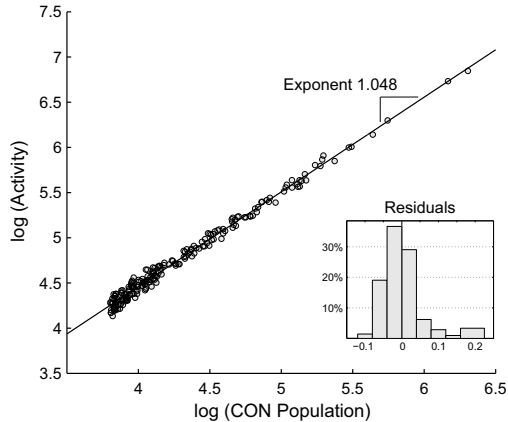}
\caption{{\bf Superlinear scaling of total spending activity with city size for the Conurbation level.}
Total spending activity is defined as the cumulative number of transactions made by city residents. Scaling exponent: $1.048$, confidence interval: [1.03,1.06], p-value: $5\cdot 10^{-204}$, $R^2=98.88\%$.}
\label{fig::scaling}
\end{figure}

\begin{figure}[h]
\centering
\includegraphics{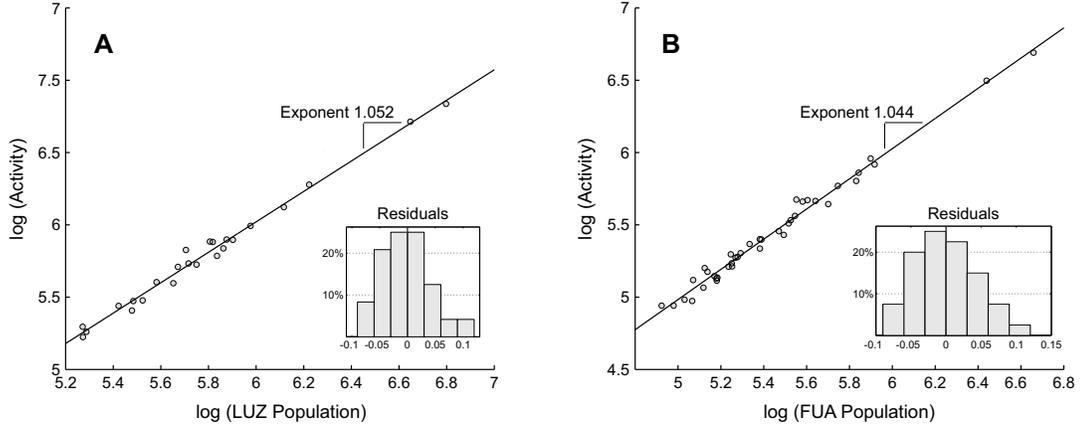}
\caption{{\bf Superlinear scaling of total spending activity with city size.}
A: For the level of Large Urban Zones. Scaling exponent:  $1.052$, CI: [1.00,1.10], p-value: $9\cdot 10^{-23}$, $R^2=98.84\%$.
B: For the level of Functional Urban Areas. Scaling exponent: $1.044$, CI: [1.00,1.08], p-value: $2\cdot 10^{-37}$, $R^2=98.67\%$.
Total spending activity is defined as the cumulative number of transactions made by city residents.}
\label{fig::scaling_fua_con}
\end{figure}

In Fig.~\ref{fig::scaling}, the total activity of each Conurbation is plotted against its size at the log-log scale. We observe a superlinear scaling with the exponent of 1.048. Statistical significance of the trend is further validated by considering the confidence interval for the exponent. Fig. \ref{fig::scaling_fua_con} confirms this finding for the two other levels of city definition (LUZs and FUAs). Importantly, the exponents for all city levels are approximately the same (around 1.05), indicating that the uncovered scaling is a distinctive feature of urban areas, regardless of the adopted definition of a city boundary.  

As we have just demonstrated, the bank card spending of individuals follows the same pattern that has been found for other socioeconomic parameters--they are boosted up in bigger cities. However, the obtained exponent of $1.05$ is lower than those obtained for other aspects of human activity (e.g., human communication scales with the exponent of $1.12$ \cite{schlapfer2012scaling}). One of the reasons might be that the considered total spending actually includes a broad range of purchases of different goods and services. Some of them fall under the common needs, which arise regardless of the place of residence (e.g., buying food), and their volume might not depend on a city size that much, while there are also more specific purchases that largely depend on the available options or might even be partially motivated by those. 

With the data at our disposal, we are also able to distinguish different target categories of businesses. 
Table~\ref{scaling_param} gives the scaling exponents for a few major business categories. To confirm the validity of the trends, we also provide confidence intervals and p-values. The fraction of activity represented by the corresponding business category is in the last column. 
As can be seen, entertaining activities like traveling, going out for a drink, dinner, or a party, are strongly boosted by city size, with a scaling exponent over 1.1. In bigger cities, people seem to engage more easily in social activities, which confirms the suggestion of \cite{schlapfer2012scaling}. Similarly, categories such as wellness, beauty, and fashion also possess higher-than-average statistically significant exponents.

\begin{table}[!ht]
\caption{
{\bf Scaling of customers' activity with CON size for different business categories.}}
\begin{tabular}{|l|c|c|c|c|}
\hline
{\bf Business category} & {\bf Exponent} & {\bf Confidence intervals} & {\bf p-value} & {\bf Fraction of activity} \\ \hline
Everyone & 1.048 & [1.033,1.061] & 3e-9 \% & 100\% \\ \hline
Bars, restaurants, and clubs & 1.122 & [1.072, 1.171] & 2.6e-6\% & 7.04\% \\ \hline
Travels & 1.108 & [1.071,1.145] & 2.6e-8\% & 6.45\% \\ \hline
Health institutions & 1.059 & [1.035, 1.084] & 3.0e-6\% & 4.75\% \\ \hline
Entertainment & 1.108 & [1.064, 1.152] & 2.1e-6\% & 0.44\% \\ \hline
Gas & 0.992 & [$0.966$,1.017] & 50.9 \% & 10.45\% \\ \hline
Supermarkets & 1.027 & [$1.009$,1.046] & 3.1e-3 \% & 29.41\% \\ \hline
Wellness, beauty and fashion & 1.078 & [$1.049$, 1.108] & 3.4e-7\% & 16.9\% \\ \hline
Others & 1.074 & [1.044, 1.104] & 2.1e-6 \% & 24.56\% \\ \hline
\end{tabular}
\label{scaling_param}
\end{table}

On the contrary, and as one could have foreseen, fundamental needs 
are less impacted by city size, with exponent values almost equal to 1. Living in a big or a small town does not seem to affect one's attendance to grocery stores, supermarkets, or gas stations. And since those activities cover 40\% of the total spending, they considerably lower the overall scaling exponent.

The other spending parameters considered in this paper demonstrate diverse behaviors in terms of scaling. Let us go over a few quantities (all others, together with the respective graphs, can be found in S1 Table and S2 Fig.). 
The average amount per transaction seems to be generally independent on the city size. Conversely, a statistically significant scaling trend appears for individual average spending diversity, suggesting that customers from larger cities have a slightly broader variety of purchases. The exponent of 0.05 together with a p-value of $2\cdot 10^{-15}\%$ for Conurbations proves the increasing trend for the average diversity with nearly 100\% confidence. The trend also seems to be statistically significant for LUZs and FUAs.
In case of mobility, the patterns are mixed. The most meaningful relation is observed for the CON level: local mobility exhibits a downward trend, with a scaling exponent of -0.031, while distant travels show a positive scaling with a noticeable exponent of 0.158. It appears that in larger cities, people are able to satisfy their shopping needs closer to their home, thus sparing them from longer journeys to perform local transactions. The uncovered trend for distant mobility (strongly increasing with population) indicates that inhabitants of large cities explore the country on a wider dimension. Nevertheless, mobility trends strongly depend on the city definition level. The mobility patterns at the scale of a Large Urban Zones are quite different from those at a core city scale.

\section*{Classification of Spanish cities beyond the impact of demography and scale}

Scaling laws described in the previous section explain how the economic behavior of city residents is expected to change with a city size. However, the actual values of spending parameters deviate from the estimations. 
For example, as can be seen from S1 Table, even when the trends are statistically significant, the portion of the observed variations they are able to explain ($R^2$) sometimes happens to be as low as $20\%$. Values higher than the corresponding trend can be treated as an overperformance of a city from the considered perspective. On the contrary, values below the trend may be interpreted as an underperformance. 
Similarly to \cite{bettencourt2010urbscaling}, we quantify those deviations as log-scale residuals (i.e., the difference between the decimal logarithm of the actual city characteristic and the decimal logarithm of the corresponding trend estimation). 
These residuals are computed in relation to size-specific estimates and can therefore be used for a qualitative comparison of different cities, regardless of their population.
Since the five urban parameters were normalized beforehand for the variability of age and gender, the residuals are also free from the impact of a demography. 
This allows the definition of a novel classification of Spanish cities, revealing the impact of local circumstances on residents' spending behavior.

Residuals of the five urban parameters represent a distinctive signature for a city. We consider each city definition level separately, which results in three separate sets of city signatures. In order to bring the residuals of different parameters on a common scale, we further normalize them by the standard deviation (while the mean is always zero by the definition of a trend).
Similarity between the signatures of cities within a given level is assessed with the k-means clustering algorithm \cite{macqueen1967kmeans}. In order to stabilize the separation of clusters, we apply the majority voting across several dozens of iterations.
According to the silhouette metric \cite{rousseeuw1987}, the most optimal solution divides cities into three clusters in cases of CONs and FUAs, and two clusters for LUZs (see S2 Text and S3 Fig. for a detailed description of the metric and received values). However, at all levels, the division into three clusters introduces an additional pattern, meaningful for the qualitative interpretation of results.
We also observe a largely consistent hierarchy between the two- and three-cluster cases, quantified as 89\% agreement for CONs, 90\% for FUAs, and 96\% for LUZs (given as proportion of cities remaining in the same cluster, under the assumption that one of the two clusters remains and the other one splits in two). Therefore, the classification into three categories of cities is retained as the basic one for the presentation of further results.

\subsection*{Spending profiles of the received categories of Spanish cities}

\begin{figure}[h!]
\vspace{0.3 cm}
\centering
\includegraphics[width=\linewidth]{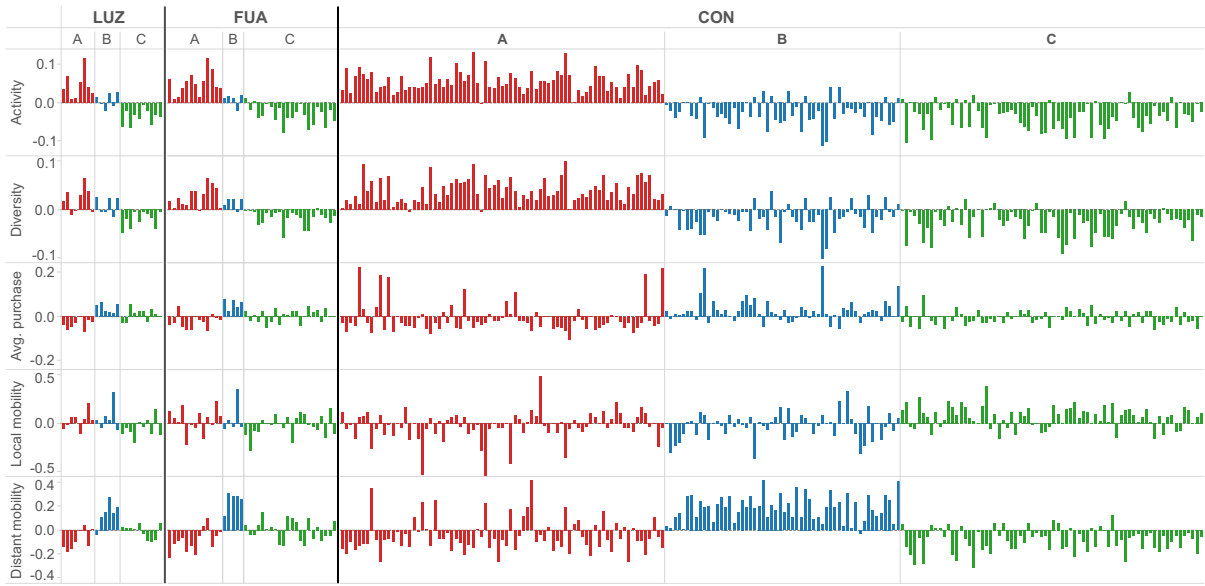}
\caption{{\bf Deviations of spending parameters from their respective scaling trends with city size, for the cities defined at the level Large Urban Zones (LUZ), Functional Urban Areas (FUA), and Conurbations (CON).}
Colors indicate three clusters of cities obtained based on the k-mean clustering in accordance with Fig.~\ref{fig::urban_clustering}}
\label{fig::residuals}
\end{figure}

Each of the received city clusters can be characterized with a distinct profile of the residents' spending behavior. Differences between particular profiles are well recognizable from the deviations of particular parameters of bank card usage (Fig.~\ref{fig::residuals}).
Major distinctions are provided by the combination of spending activity and diversity, as well as distant mobility. The first two parameters, which are in any case correlated as higher number of transactions implies higher diversity (see also S4 Fig., which shows the mutual correlation between the five spending parameters and gives another visual perspective on the distinction of the clusters), explain the separation of the cluster A (red). The variations of distant mobility justify the further split of clusters B (blue) and C (green). Importantly, those distinctions are consistent across the city levels.

\begin{figure}[h!]
\centering
\includegraphics[width=\linewidth]{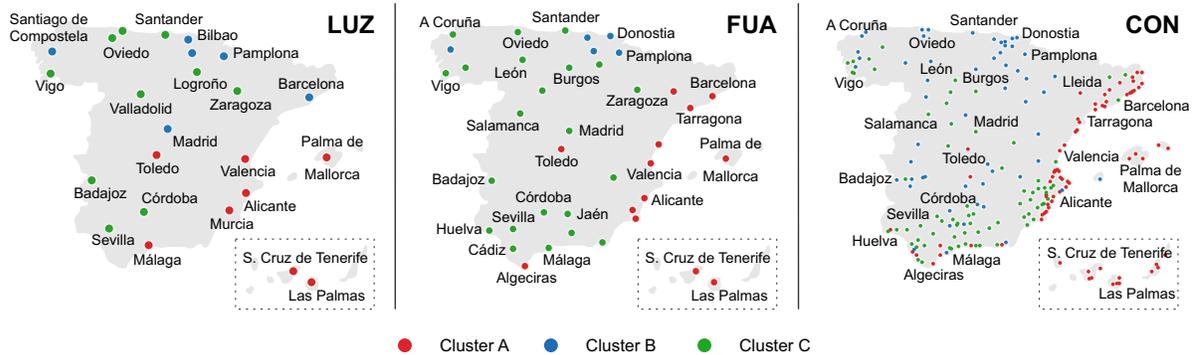}
\caption{{\bf Classification of Spanish cities into three categories based on the spending behavior of their residents.}
Classification was performed separately for the three different city definition levels--Large Urban Zones (LUZ), Functional Urban Areas (FUA), and Conurbations (CON). Clustering into two categories can be, to large extent, recreated by merging clusters B and C into one (the consistency between the two- and three-cluster cases was quantified as high as $95\%$ for LUZs, $90\%$ for FUAs, and $89\%$ for CONs).}
\label{fig::urban_clustering}
\end{figure}

Received clusters can be further interpreted based on their spatial alignment (Fig.~\ref{fig::urban_clustering}), which exhibits a high robustness across the three city definition levels. The variations can be attributed, to a large degree, to the changes of sample size and spatial scale (e.g., different mobility patterns within LUZs and CONs). In general, a good agreement across city levels is observed and can serve as an additional evidence of consistency and stability of the approach. Below we present a detailed interpretation of the three received city clusters. 

The most distinctive city category, especially for CONs and FUAs, is the red cluster covering cities located along the most visited part of the Mediterranean coast and on the islands. This pattern clearly refers to the most touristic parts of Spain, which is further supported by the incorporation of Toledo, a World Heritage Site by UNESCO. The appearance of this ``touristic'' cluster is interesting, as our procedure relies exclusively on the economic activity of city residents.
It may indicate that inhabitants of this type of cities have a distinctive economic behavior. 
The spending profile of the red cluster is characterized by an intensified spending activity and diversity, which are accompanied by an underperformance in terms of the average purchase. This indicates that residents use their bank cards more often than for the occasional shopping, covering all types of small, everyday purchases. Negative deviations recorded for the distant mobility parameter are well understandable from the perspective of the geographic distribution of cities, especially for a group of isolated island cities.

The core of the blue cluster includes cities from Basque and Navarra regions and Santiago de Compostela. In case of LUZs, it also covers Madrid and Barcelona. In case of CONs, it expands into a much wider area in the north as well as part of the Spanish interior. 
Cities grouped within this cluster exhibit negative residuals for the activity and diversity parameters, while the residuals of the average purchase tend to be positive. The latter goes in line with the economic profile of the core cities in the blue cluster--wealthy industrial cities in the northern part of Spain, such as  Bilbao or Pamplona. 
Higher purchase power may in turn be the reason for the distant transaction values, which are much higher than expected from the scaling trend. However, these proofs of larger spending potential are combined with the decreased intensity of card usage. This points to a conclusion opposite to that for the red cluster--bank cards are used for substantial amounts of money but quite rarely.

Green cluster concentrates around the southern cities and the remaining ones from the north. 
Similar as the blue cluster, it is characterized by the negative residuals for the activity and diversity parameters. Deviations of the remaining three parameters are of a different type.
The average purchase values are close to the baseline provided by the general trend. At the same time, distant mobility mainly records negative deviations, especially at the level of CONs. Accompanied by a larger local mobility for CONs, this observation suggests an economic activity of residents concentrated around their cities, which are satisfying most of their needs. It agrees well with the fact that the green cluster consists of many service-oriented cities, serving as administrative capitals for big territories, such as Valladolid, Sevilla, and Zaragoza.

\subsection*{Socioeconomic profiles of the clusters}
 Three categories of Spanish cities have been received solely based on the individual economic activity of their residents. In this section, we examine how well these clusters correspond with the standard socioeconomic statistics. We picked three major urban indicators, available for the city levels of LUZ and FUA from the Urban Audit Survey~\cite{eur_uaudit}. These are: Gross Domestic Product (GDP, estimated based on the province quantities), unemployment (in absolute numbers), and total disposal annual income of households (further referred to as income). 
Data at the CON level were not available to the authors. As already mentioned, socioeconomic indicators are proven to scale superlinearly with city size \cite{bettencourt2007growth, bettencourt2013origins}. We confirmed that this property holds also for our socioeconomic data, with the exception of the income at the level of FUA, where the relation with city population is rather linear. 
Therefore, we construct socioeconomic metrics in the same way as our five spending parameters, based on the log deviations from the scaling estimates. 

Thorough linear correlation can be observed only for the average value of a purchase, especially at the level of LUZ. The correlation is negative with unemployment ($R^2$ = 0.84 / LUZ and 0.67 / FUA) and positive with income ($R^2$ = 0.58 / LUZ and 0.39 / FUA) and GDP ($R^2$ = 0.5 / LUZ 0.38 / FUA). Weaker, though still visible, correlations can further be noted between distant mobility and unemployment (negative) as well as distant mobility and GDP (positive). Other spending parameters do not indicate a direct linear correlation with socioeconomic statistics.

\begin{figure}[!h]
\vspace{0.5 cm}
\centering
\includegraphics[width=\linewidth]{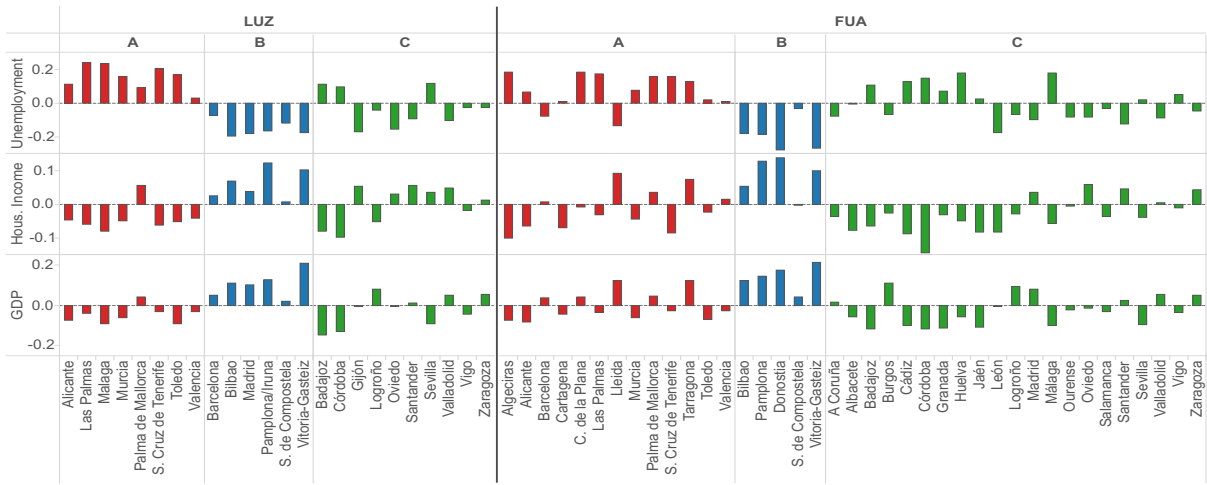}
\caption{{\bf Deviations of standard socioeconomic statistics within the three clusters of Spanish cities at the level of LUZ, FUA, and CON.}
Unemployment, Disposal Annual Income of Households, and Gross Domestic Product are quantified as log residuals from their respective scaling trends with city size. Colors correspond to the three clusters of cities obtained based on the spending behavior of city residents, as  presented in Fig.~\ref{fig::urban_clustering}.}
\label{fig::socioeconomic}
\end{figure}

Nevertheless, interesting patterns pop up when crossing the deviations of socioeconomic metrics with the three received city clusters (Fig.~\ref{fig::socioeconomic}). The most distinctive observation concerns cluster B (blue), where all cities, except Santiago de Compostela, which is very close to the trend line, exhibit high positive residuals for income and GDP and negative residuals for unemployment. It confirms a good socioeconomic condition among the cities of the blue cluster and well agrees with their spending profile--higher than expected average purchase values and distant mobility (in this case, one would assume that people demonstrate high business activity that requires distant travels or are wealthy enough to travel for leisure). The situation seems to be just the opposite for the A cluster (red). The majority of cities record high residuals for unemployment and relative underperformance in terms of income and GDP, which may indicate both social and economic problems at the city level. Individually, these go hand in hand with the decreased average purchase values and distant mobility but do not prevent a high level of general spending activity. Interpretation for cluster C (green), characterized by the mix of residuals' values, does not seem to be as straightforward as for the previous ones; however, we may observe a slight tendency to underperform in economic metrics. This agrees with the previously reported low levels of spending activity and diversity, as well as distant mobility. All observations on the correspondence of the three city clusters and socioeconomic statistics are generally consistent across the city levels, which further confirms their validity.

\subsection*{Attractiveness of the clusters to foreign visitors}
\begin{figure}[h]
\vspace{0.5 cm}
\centering
\includegraphics[width=\linewidth]{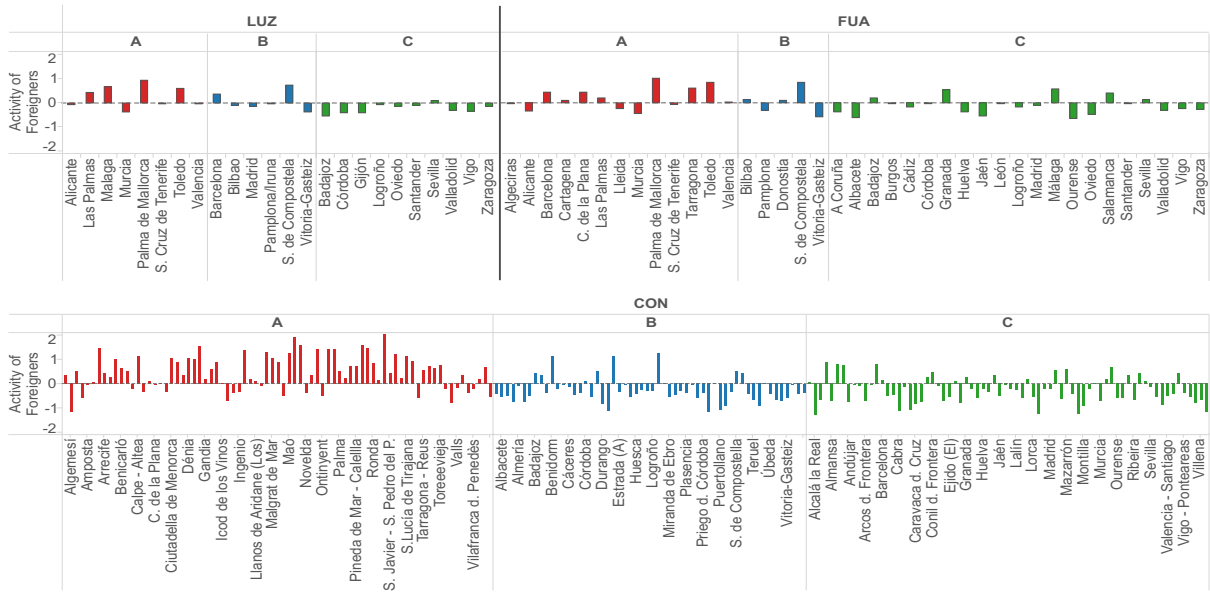}
\caption{{\bf Correspondence of the received clusters of Spanish cities with the attractiveness of these cities.}
Attractiveness is calculated based on the spending activity of foreign visitors. Presented deviations are quantified as log residuals from their respective scaling trends with city size at the levels of  LUZs, FUAs, and CONs. Colors correspond to the three clusters of cities obtained based on the spending behavior of city residents, as  presented in Fig.~\ref{fig::urban_clustering}.}
\label{fig::for_attractivenes}
\end{figure}

We have already mentioned that the distribution of cities among the clusters leads to the inclusion of the most touristic parts of Spain in cluster A, even though the grouping was performed solely based on the activity of city residents without taking visitors into account.
To explore this observation in a more quantitative manner, we compared our clusters with the attractiveness of the Spanish cities to the foreign visitors.
As proposed in \cite{sobolevsky2014mining, sobolevsky2015attract},
the attractiveness of a city can be captured e.g., with the spending activity of its foreign visitors, recorded with the point of sale transactions. Such measure exhibits a strong superlinear scaling with city size, with an exponent of around 1.5 \cite{sobolevsky2015attract}. 
Therefore, residuals from the scaling trends at each city level can be compared with the distribution of cities among the clusters, given the availability of the appropriate anonymized data containing the bank card purchases of the foreigners visiting Spain. As we can observe in Fig.~\ref {fig::for_attractivenes}, cities gathered in cluster A are indeed characterized by the spending activity of foreigners way above the trend line, while the majority of the cities from the remaining clusters tend to underperform on this measure.
The pattern is especially evident at the level of Conurbations, but to a minor extent, it is visible also for LUZs and FUAs.
In general, the correspondence of visitor attractiveness and spending habits of city residents is an interesting observation, which shows a great extent to which a touristic profile of a city affects the individual life of its inhabitants. At the same time, it also puts the proposed classification in a much broader applicability context going beyond the original idea of the impact of cities on the economic behavior of their residents.

\section*{Conclusions}
In the present study, we explored the impact of different factors such as age, gender, and place of residence on customer spending behavior, quantified by means of five different characteristics: customer activity, average value of transaction, spending diversity, local, and distant mobility. 
We found that age and gender have a major impact on spending behavior, affecting all of the above parameters. Consistent trends were obtained when correlating them with age, and the curves for genders were different from one another, similar in shape but not in absolute terms. For instance, the average purchase amount demonstrates a surprisingly steady linear increase with customer age. This phenomenon might be interpreted as a general tendency to group purchases when growing old. 
Also, the spending diversity, after a certain increase until a peak around $\approx$30 years of age (consistent with the average age of first childbirth), starts a steady and nearly linear decrease. In general, confirmed impact of the customers' demographics on their bank card activity implies the necessity to compensate for such factors while building and comparing aggregated patterns, e.g., spending profiles of different urban areas. 

Next, we found that the size of a city of residence has a noticeable impact on all the characteristics of individual behavior, except for the average amount of purchases, in a way that often might be described as a statistically significant power law scaling. The overall spending activity scales superlinearly with city size--a fact that goes in line with the previous findings regarding other socioeconomic urban quantities. At the same time, individual spending in different types of businesses shows substantially different scaling behavior, which hints for a noticeable shift in the categories of customer activity within the cities of a different scale.

However, each city posses its own unique characteristics in terms of the examined spending parameters, which go beyond the existing trends of a general superlinear scaling. We demonstrated that the deviations from the trends' baselines can be regarded as distinct signatures of cities, forming a solid basis for a scale-free comparison and classification of urban areas. The approach was tested on Spanish cities defined at three different scales--Large Urban Zones, Functional Urban Areas, and Conurbations. This purely data-driven classification, independent from any spatial or topological considerations, revealed three meaningful categories of Spanish cities. The categories have shown to be distinct in terms of the spending patterns of their residents, geographic alignment, and the standard socioeconomic statistics derived with the external metrics of GDP, unemployment, and income. Furthermore, obtained results were found to capture meaningful economic patterns beyond the scope of the considered data--certain categories, while being derived solely based on the behavior of city residents, corresponded well with the attractiveness of those cities to foreign visitors. This fact validates the applied merit of the approach. 

As a final remark, we should point out that the proposed classification remained, to a certain extent, stable for the different city definitions, which constitutes additional evidence of its robustness. Moreover, although in the paper the approach is applied for the case of Spain, we believe it to be widely applicable to any other country, provided that the appropriate data are available.

\section*{Acknowledgments}
The authors would like to thank BBVA for providing the dataset for this research. Special thanks to Assaf Biderman, Marco Bressan, Elena Alfaro Martinez and Mar\'ia Hern\'andez Rubio for organizational
support of the project and stimulating discussions. 
We further thank the BBVA, MIT SMART Program, Center for Complex Engineering Systems (CCES) at KACST and MIT, the National Science Foundation,the MIT Portugal Program, the AT\&T Foundation, The Coca Cola Company, Ericsson, Expo 2015, Ferrovial,
The Regional Municipality of Wood Buffalo, Volkswagen Electronics Research Lab and all the members of the MIT Senseable City Lab Consortium for supporting the research. IS and BH acknowledge funding from the Austrian Science Fund (FWF) through the Doctoral College GIScience (DK W 1237-N23).

\clearpage

\begin{flushleft}
\bibliography{bbva_arxiv}
\end{flushleft}

\clearpage

\section*{Supporting Information}
\bigskip
\subsection*{S1 Text. Demographic normalization}
Let us describe the normalization procedure used to account for demographic discrepancies between cities. The idea is to compare the observed value of a given parameter with its theoretical expected one (computed using the city demographic profile). Let $(p_c)_{c \in C}$ be the measured parameter where $C$ denotes the entire set of customers, $C_X$ the subset containing only customers from city X and $C_{g, a}$ the customers of gender g and age a.
The average quantity for a given gender g and age a and for a given city X are 
\begin{equation}
Q_{g,a} = \frac{ \displaystyle{\sum_{c \in C_{g, a}} p_c}}{|C_{g,a}|} \hspace{1cm} Q_{X} = \frac{ \displaystyle{\sum_{c \in C_{X}} p_c}}{|C_{X}|}\nonumber  
\end{equation}
The expected value of the parameter based on the demography of city X is
\begin{equation}
E_{X} = \frac{ \displaystyle{\sum_{g, a }} |C_{X} \cap C_{g, a}| Q_{g, a}}{|C_{X}|}  \nonumber  
\end{equation}
In the end, the normalized value used as a measure of city X economic behavior is $\frac{Q_X}{E_X}$
\\

\subsection*{S2 Text. Validation of the clustering scheme}

Similarity between the signatures of cities within each definition level i.e., CONs, FUAs and LUZs, was assessed with the k-means clustering algorithm \cite{macqueen1967kmeans}. In order to select the optimal number of clusters, we validated different approaches with the silhouette metric \emph{s(i)} \cite{rousseeuw1987}, with \emph{k} $\in$ [2;10]. Silhouette aims to reflect how well each object fits to its cluster based on the comparison of an object dissimilarity (in our case, the Euclidean distance) to the points grouped in the same cluster and to the points grouped within the next best fitting cluster. It is computed according to the equation: 

\begin{equation}
s(i) = \frac{ b(i) - a(i)}{max\{a(i), b(i)\}},  \nonumber  
\end{equation}
where \emph {a(i)} is the average dissimilarity of a city \emph {i} to the other cities assigned to the same cluster, and \emph {b(i)} is the average dissimilarity of a  city \emph {i} to its next best fitting cluster. Silhouette varies in the range of [-1;1]. Positive values indicate a good match with the own cluster (small \emph {a(i)}) and a bad match with the neighboring cluster (high  \emph {b(i)}). On the contrary, negative values indicate that a data points is more similar to the neighboring cluster, while values around 0 imply that a point is on the edge of two clusters. 
In our case, we compared the average values of \emph{s(i)} across the clustering schemes with different \emph{k}, assuming that the highest values indicate the optimal split of cities. Received values are presented in the Figure S3. We observed that the silhouette metric peaked at \emph{k} = 3 for the levels of CONs and FUAs, and \emph{k} = 2 for LUZs. For the sake of consistency, we selected three-cluster approach for all the levels. Additionally, we validated selected algorithm, i.e. k-means, against its common variation, that is k-medoids \cite{kmedoids}. As the latter one resulted in lower silhouette values for all tested \emph{k}, we retained the k-means approach. 

\pagebreak
\subsection*{S1 Table. Scaling of the five characteristics of individual spending behavior with city size.}
\begin{table}[!ht]
\centering
\begin{tabular}{|c|c|c|c|c|c|}
\hline
{\bf Parameter} & {\bf City definition} & {\bf Exponen} & {\bf Confidence intervals} & {\bf p-value} & {\bf $R^2$} \\ 
\hline
\multirow{3}{*}{Activity} & LUZ & 1.052 & [1.0,1.1] & 9e-21 \% & 98.84\% \\
\cline{2-6}
& FUA & 1.044 & [1.0,1.08] & 2e-35 \% & 98.67\% \\
\cline{2-6}
& CON & 1.048 & [1.03,1.06] & 5e-202 \% & 98.88\% \\
\hline
\multirow{3}{*}{Avg. amount } & LUZ & -0.007 & [-0.05,0.03] & 71.2 \% & 0.6\% \\
\cline{2-6}
&  FUA & 0.002 & [-0.03,0.04] & 87.9 \% & 6e-4 \% \\
\cline{2-6}
& CON & 0.008 & [-0.007,0.02] & 28.6 \% & 0.5\% \\
\hline
\multirow{3}{*}{Diversity} & LUZ & 0.033 & [0.0,0.064] & 4.2 \% & 17.5\% \\
\cline{2-6}
 & FUA & 0.035 & [0.01,0.06] & 0.49 \% & 19.0\% \\
\cline{2-6}
& CON & 0.051 & [0.04,0.06] & 2e-15 \% & 26.1\% \\
\hline
\multirow{3}{*}{Distant mob.} & LUZ & -0.06 & [-0.24,0.11] & 45.8 \% & 2.5\% \\
\cline{2-6}
& FUA & 0.035 & [-0.1,0.16] & 60 \% & 0.7\% \\
\cline{2-6}
& CON & 0.158 & [0.11,0.20] & 6e-11 \% & 18.65\% \\
\hline
\multirow{3}{*}{Local mob.} & LUZ & -0.10 & [-0.24,0.04] & 15.2 \% & 9.1\% \\
\cline{2-6}
& FUA & -0.073 & [-0.17,0.03] & 16.1 \% & 5.1\% \\
\cline{2-6}
& CON & -0.031 & [-0.07,0.01] & 15.4 \% & 1.0\% \\
\hline
\end{tabular}
\label{scaling_param}
\end{table}

\bigskip
\subsection*{S1 Fig. Impact of age and gender on the total amount of money spent by BBVA customers in 2011.}

\begin{figure}[h!]
\centering
\includegraphics{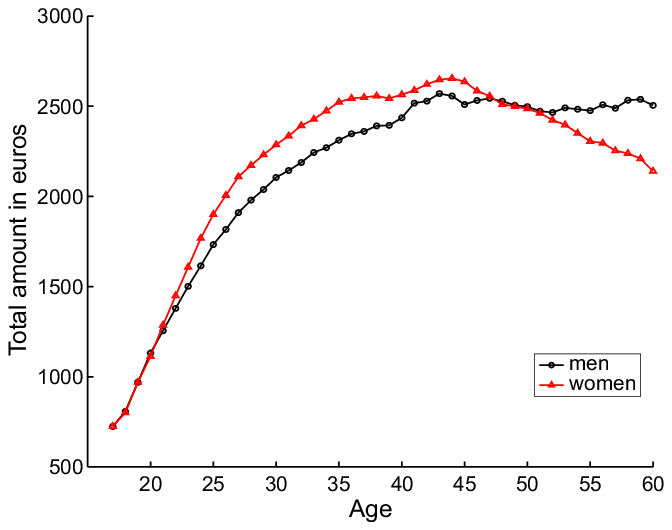}
\end{figure}

\clearpage
\subsection*{S2 Fig. Scaling of spending diversity with city size.}
As mentioned in the main text, the spending diversity exhibits a small but consistent scaling with city size for the three definitions considered: Large Urban Zones (LUZs), Functional Urban Areas (FUA), and Conurbations (CON).
{\small LUZ exponent: $3.3\%$, CI: [0.0,0.064], p-value: $4.2 \%$, $R^2 = 17.5 \%$. FUA exponent: $3.45\%$, CI: [0.01,0.06], p-value: $0.49 \%$, $R^2 = 19.0\%$. CON exponent: $5.1\%$, CI: [0.04,0.06], p-value: $2e$-$15 \%$, $R^2 = 26.1\%$.}

\begin{figure}[h!]
\centering
\includegraphics[width=0.86\linewidth]{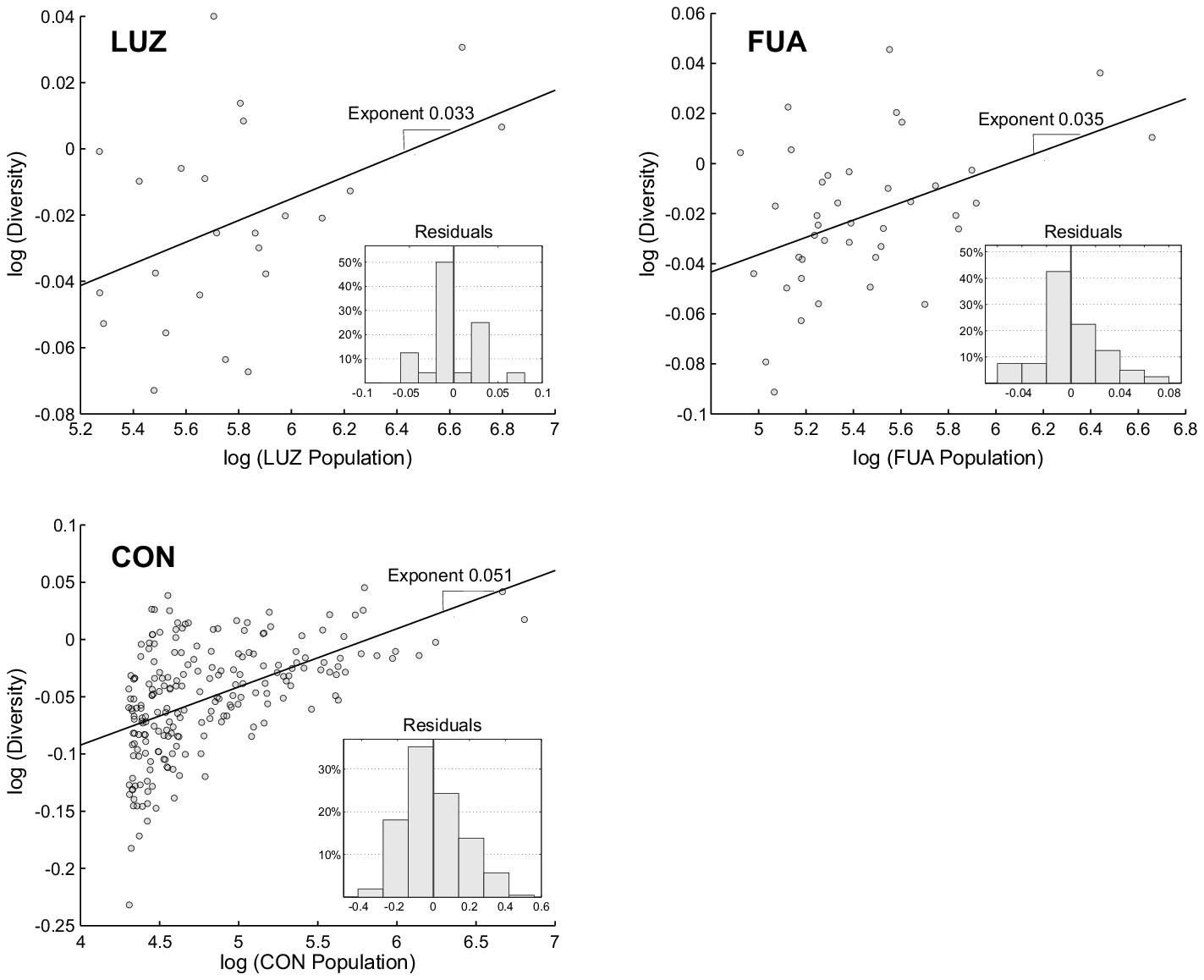}
\end{figure}

\subsection*{S3 Fig. Values of the silhouette metric for different k-mean clustering schemes.}
We varied $k$ between 2 and 10, separately for the three levels of city definition: Large Urban Zones (LUZs), Functional Urban Areas (FUA), and Conurbations (CON). Higher values indicate a better fit of data points to the clusters they were assigned (more appropriate clustering approach).

 \begin{figure}[h!]
\centering
\includegraphics{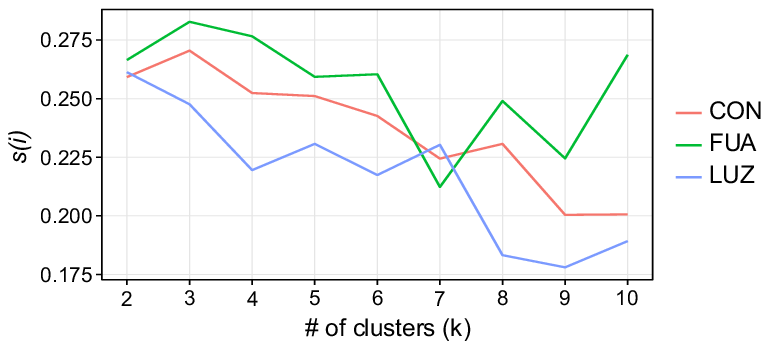}
\end{figure}

\clearpage
\subsection*{S4 Fig. Correlations between the five characteristics of customers' spending behavior.}
In order to examine how different spending characteristics relates to one another in terms of city  under- or overperformance , we visualized mutual correlation plots of their residuals, for the pairs of all characteristics, at each of the three city definitions, LUZs, FUAs, and CONs. 
Data points of the cities from different clusters are plotted with the colors of the corresponding clusters. Only activity and diversity happen to be strongly correlated for all the three levels of city definition.
Certain dependencies could be noticed between activity and the average amount of purchase; however, those dependencies are opposite for the FUA level on one hand and LUZ and CON levels on the other. Also, average purchase and distant mobility show a weak correlation but only at LUZ and FUA levels. All other pairs of characteristics seem to be independent from one another. At the same time, clusters seem to be quite distinctive at the majority of the plots.

 \begin{figure}[h!]
\centering
\includegraphics[width=0.86\linewidth]{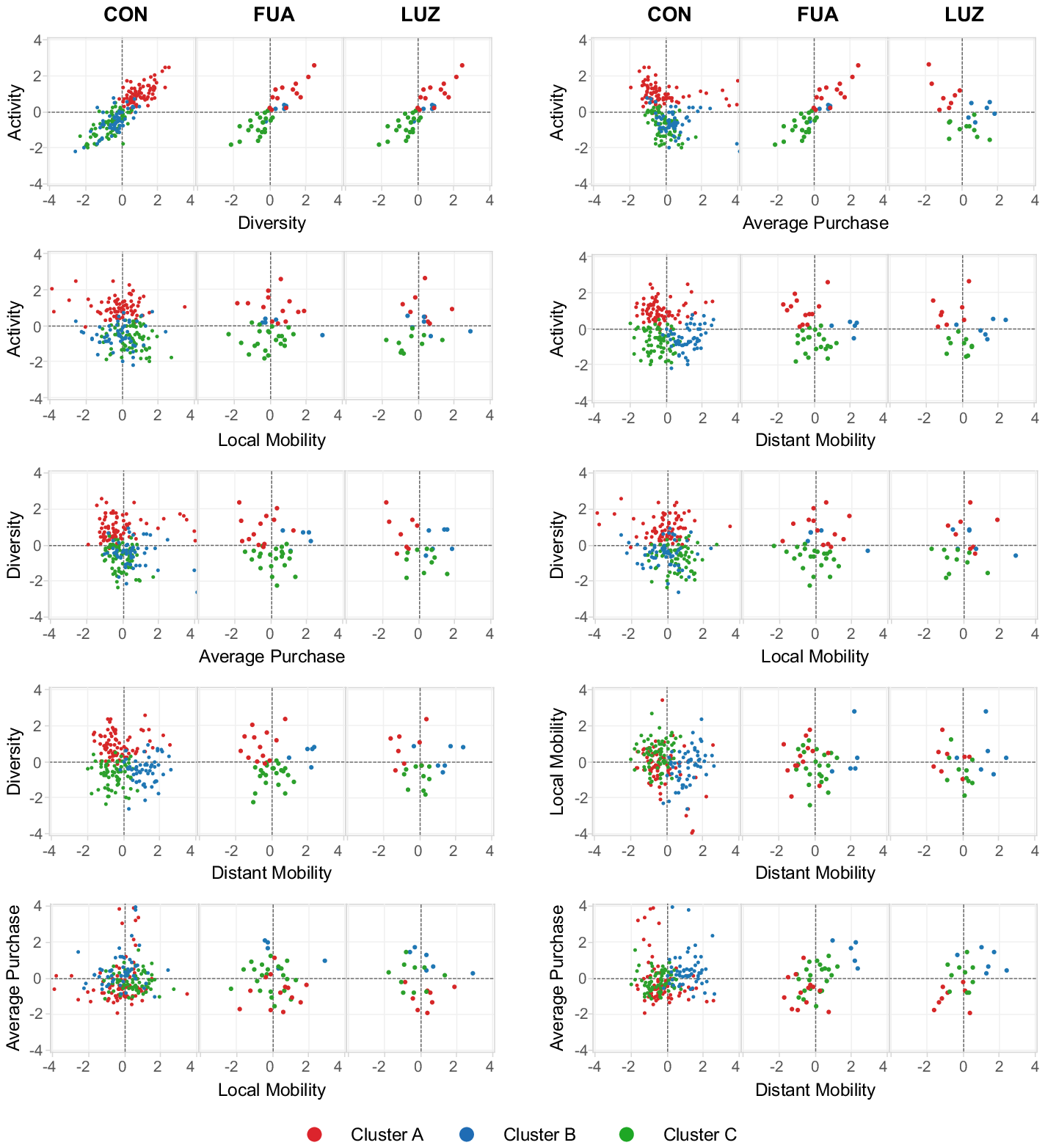}
\end{figure}

\end{document}